\documentclass{ws-mpla}
\usepackage[super]{cite}
\usepackage{graphicx}
\usepackage{epsfig}
\usepackage[normalem]{ulem}
\newcommand{\cmb}{\textrm{\tiny CMB}}

\begin{document}

\markboth{JUAN C. BUENO SANCHEZ}
{ON THE BREAKING OF STATISTICAL ISOTROPY THROUGH INFLATIONARY RELICS}

\catchline{}{}{}{}{}

\title{ON THE BREAKING OF STATISTICAL ISOTROPY THROUGH INFLATIONARY RELICS}
\author{\footnotesize JUAN C. BUENO S\'ANCHEZ}
\address{Centro de Investigaciones en Ciencias  B\'asicas y Aplicadas, Universidad Antonio Nari\~no\\Cra 3 Este \# 47A-15, Bogot\'a D.C. 110231, Colombia.\\Departamento de F\'isica, Universidad del Valle\\Calle 13 No 100-00, Santiago de Cali A.A. 25360, Colombia.\\Escuela de F\'isica, Universidad Industrial de Santander\\Ciudad Universitaria, Cra. 27 Calle 9, Bucaramanga 680002, Colombia.\\juan.c.bueno@correounivalle.edu.co}

\maketitle

\begin{abstract}
In this talk we elaborate on a mechanism to generate local contributions to the curvature perturbation in isolated patches of the Cosmic Microwave Background (CMB). The mechanism, based on the generation of an out-of-equilibrium configuration in fluctuating scalar fields of mass $m\sim H$ during a sustained stage of fast-roll inflation, has been recently shown to be capable of accounting for some of the most robust large-angle anomalies detected in the CMB. In this talk, we show in detail how the embedding of the mechanism into models including vector fields can result in the breaking of statistical isotropy in isolated patches of the CMB. 
 
\end{abstract}

\ccode{PACS Nos.: include PACS Nos.}

\section{Introduction}	
Thanks to a wealth of high precision observations of the CMB \cite{Bennett12,Hinshaw:2012aka,Ade:2015xua,Ade:2015lrj}, cosmological inflation, and in particular single-field inflation, is widely accepted as the simplest paradigm to explain the origin of the observed Universe. However, cosmological inflation still faces a number of difficulties, the most obvious one being the large class of models which, while consistent with data, have very different implications for particle physics. Another less pressing difficulty is the persistence, for more than a decade now, of large-angle anomalies in the CMB, which suggests that single-field inflation might need an extension of some kind. These anomalies, entailing the breaking of statistical homogeneity and isotropy of the CMB, were observed for the first time by the WMAP satellite \cite{Bennett:2003bz,Bennett:2010jb} and later confirmed by Planck \cite{Ade:2013nlj,Ade:2015hxq}. Since their existence seems to pose a relative challenge for single-field inflation, an important theoretical effort has been dedicated over the past decade to elucidate their origin. 

\section{A mechanism to generate statistical inhomogeneity}
Here we make use of a framework recently discussed in the literature aiming to shed light on the origin of the CMB anomalies by attributing their origin to the statistical inhomogeneity developed by different isocurvature fields present during inflation. Following \cite{Sanchez:2014cya,Sanchez:2016lfw}, we consider a system of spectator scalar fields $\sigma$ and $\chi$, minimally coupled to gravity and interacting through a term of the form\footnote{Such an interaction term has been widely considered in a number of applications, as in (p)reheating \cite{Dolgov:1982th,Abbott:1982hn,Traschen:1990sw,Dolgov:1989us,Kofman:1994rk,Shtanov:1994ce,Kofman:1997yn,Felder:1998vq}, moduli trapping \cite{Kofman:2004yc,Watson:2004aq}, inflation model building \cite{Kadota:2003tn,BuenoSanchez:2006eq,BuenoSanchez:2006ah,Green:2009ds}, in the generation of non-Gaussianity \cite{Barnaby:2009mc,Wu:2006xp,Lee:2011fj} and in the evolution of coupled flat directions \cite{Enqvist:2011pt,Sanchez:2012tk}.} $g^2\sigma^2\chi^2$. Ignoring the interactions of these fields with other degrees of freedom, the Lagrangian of our system is
\begin{equation}\label{eq10}
{\cal L}=\frac12\partial_\mu\sigma\partial^\mu\sigma-V(\sigma)+\frac12\partial_\mu\chi\partial^\mu\chi
-\frac12 m_{0\chi}^2\chi^2-\frac12\,g^2\sigma^2\chi^2\,,
\end{equation}
where $g$ is a coupling constant and $m_{0\chi}$ is the bare mass of $\chi$. The scalar potential for $\sigma$ is $V(\sigma)=\frac12\,m_\sigma^2\sigma^2$, taking $m_\sigma^2=c_\sigma H^2$ where $c_\sigma$ is a constant.

The prototype isocurvature field $\sigma$ here considered is allowed to have a large mass, $m\sim H$, and hence we focus on $c_\sigma={\cal O}(10^{-1})$ \cite{Sanchez:2016lfw}. The benefit from this choice is that the existence of fields with such a large mass is generic in supergravity theories \cite{Copeland:1994vg,Dine:1995uk}. However, using this fields also generates a problem for the initial conditions. As shown in \cite{Sanchez:2016lfw}, in order to develop statistical inhomogeneity the field $\sigma$ must begin the slow-roll with a large value compared to $H$, which can be achieved through a sustained stage of fast-roll inflation.

\subsection{A primary fast-roll stage}
To describe the evolution of the classical field $\sigma$ during inflation we must specify the different phases of inflation. We thus write the total length of inflation as
\begin{equation}
N_{\rm tot}=N_{\rm fr}+N_{\rm sr}\,\,,\,\,N_{\rm sr}=N_{\rm sr}^p+N_*\,,
\end{equation}
where ``fr'' and ``sr'' stand for fast-roll and slow-roll, respectively. We allow the primary phase of inflation to contain a slow-roll stage, where $N_{\rm sr}^p>0$ denotes its length. Also, the largest cosmological scales cross outside the horizon $N_*$ $e$-foldings before the end of inflation. Observations typically demand that $40\lesssim N_*\lesssim 70$.

If we stick to a constant value of $\epsilon$, the Hubble parameter is $H\propto a(t)^{-\epsilon}$, where  $a(t)=(1+\epsilon H_0t)^{1/\epsilon}$. Using this, we can solve the perturbation modes equation
\begin{equation}
\delta\ddot\sigma_k+3H\dot\delta\sigma_k+\left(\frac{k^2}{a^2}+c_\sigma H^2\right)
\delta\sigma_k=0\,.
\end{equation}
Solving for the above using the Bunch-Davies solution in the subhorizon limit $k/aH\to0$ we find the perturbation spectrum 
\begin{equation}
{\cal P}_{\delta\sigma}(k)\equiv\lim_{k/aH\to0}\frac{k^3|\delta\sigma_k|^2}{2\pi^2}=\gamma\,\frac{H^2}{4\pi^2}\left(\frac{k}{aH}\right)^{3-2\nu}\,,
\end{equation}
where $\gamma\equiv\frac{2^{-1+2\nu}\Gamma(\nu)^2}{\pi(1-\epsilon)^{1-2\nu}}$. If we assume that the field vanishes at the beginning of inflation, the field variance $\Sigma^2\equiv\langle(\sigma-\bar\sigma)^2\rangle$ is
\begin{equation}
\Sigma^2(N,c_\sigma,\epsilon)=\gamma\frac{H^2}{4\pi^2(3-2\nu)}\left(1-e^{-(3-2\nu)N}\right)\,,
\end{equation}
where $N$ is the number of elapsed $e$-foldings from the beginning of inflation.
\begin{figure}[htbp]
\centerline{\includegraphics[width=3.0in]{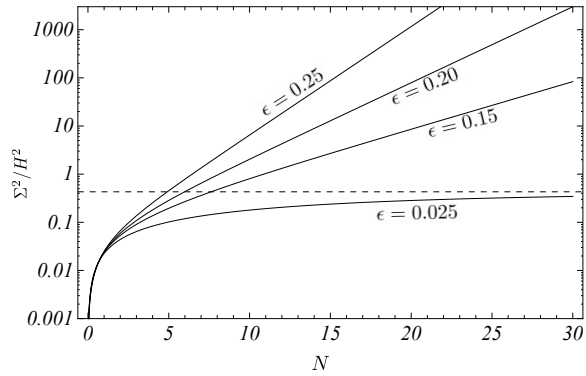}}\caption{Evolution of $\Sigma^2/H^2$ during inflation (using different $\epsilon$) for a field with mass $c_\sigma=0.15$.}\label{fig1}
\end{figure}

In Fig.~\ref{fig1} we plot the evolution of $\Sigma^2/H^2$ during inflation, using different values of $\epsilon$. Our plot shows that for a sufficiently large $\epsilon$, the ratio $\Sigma^2/H^2$ grows unbounded due to the rapid scaling of $H$. However, if $\epsilon$ is small (curve with $\epsilon=0.025$), the growth of $\Sigma^2/H^2$ becomes bounded (dashed line), and hence typical values of $\sigma$ always remain in the same order of magnitude as $H$.

\subsection{Trapping the rolling $\sigma$}
Once the fast-roll stage is over and typical field values are $\sigma>\sigma_c$, the field starts to slow-roll. Using the slow-roll approximation we write
\begin{equation}\label{eq18}
\sigma(t)=\sigma_0\,\exp(-c_\sigma N/3)\,,
\end{equation}
where $N$ counts the number of $e$-foldings after all CMB scales cross outside the horizon. Eventually, it reaches the critical value $\sigma_c\sim g^{-1}H$ and the $\chi$ field starts undergoing particle production during inflation. This production of particles triggers a trapping mechanism that forces $\sigma$ to oscillate about the origin of its effective potential. In this oscillatory regime, the typical amplitude of the field scales as $\sigma\propto a^{-3/2}$ and hence it becomes negligible by the end of inflation. However, since due to the inflationary fluctuations of $\sigma$ the trapping occurs in different places at different times, it becomes possible that in some regions (called out-of-equilibrium patches) $\sigma$ does not become trapped before the end of inflation, thus retaining a large, out-of-equilibrium value that can be considered as a relic from a primary stage of fast-roll inflation. Consequently, the role played by the interaction term is to motivate the existence of regions in the observable Universe where $\sigma$, thanks to its retaining a large value, can leave an observable imprint in the CMB. 

In order to describe the evolution of the field from the slow-roll to the oscillatory stage, we must solve the coupled system of equations arising from (\ref{eq10}). In the Hartree approximation, the dynamics of $\sigma$ and $\chi$ is determined by the equations
\begin{equation}\label{eq11}
\ddot\sigma+3H\dot \sigma+(c_\sigma H^2+g^2\langle\chi^2\rangle)\sigma=0\,,
\end{equation}
\begin{equation}\label{eq12}
\ddot\chi_k+3H\dot\chi_k+\left(\frac{k^2}{a^2}+g^2\sigma^2\right)\chi_k=0\,,
\end{equation}
where 
\begin{equation}\label{eq13}
\langle\chi^2\rangle=\frac1{(2\pi)^3}\int|\chi_k|^2d^3k\,.
\end{equation}
We solve this system numerically, using an initial condition $\sigma>\sigma_c$ motivated by the hypothesized existence of a sustained fast-roll stage. In our numerical solution\footnote{We solve the coupled system in time steps $\Delta t=0.025H_*^{-1}$ so that a total of 40 modes cross outside the horizon per $e$-folding, thus contributing to $\langle\chi^2\rangle$ in Eq.~(\ref{eq13}). We integrate the coupled system for seven $e$-foldings, thus employing modes with comoving wavenumber from $k=H_*$ (crossing outside the horizon at $t_c$) up to $k=e^7H_*$. Moreover, all of the modes $\chi_k$ considered in the solution begin the evolution in the Bunch-Davies vacuum.} we allow the production of the $\chi$ field only when $\sigma(t_c)=\sigma_c$. Using different multiplicities for the $\chi$ field, with $N_\chi={\cal O}(10^2)$, we find that the transition to the oscillation regime of $\sigma$ takes a few $e$-foldings to complete. 

\section{Shedding light on the origin of CMB anomalies}
Recently, it has been investigated the possibility that CMB anomalies are due to the statistical inhomogeneity obtained by different isocurvature fields of mass $m\sim H$ present during inflation \cite{Sanchez:2014cya,Sanchez:2016lfw}. The framework presented shows that if primary inflation contains a sustained stage of fast-roll, like the one studied in the previous section, and primary slow-roll lasts only for a few tens of $e$-foldings at most, then it is feasible that the large expectation value obtained by the isocurvature fields is responsible for some of the CMB anomalies. In particular, in Ref.~\refcite{Sanchez:2016lfw} it is shows that making use of the inhomogeneous reheating and of the curvaton mechanism, it is possible to account for the Cold Spot, the power deficit at low $\ell$ and the breaking of statistical isotropy. In this talk, I focus on the latter and show that the curvature perturbation contributed by the vector field, using the vector curvaton mechanism, indeed obtains a patchy structure, being observable in isolated patches of the CMB. 

\subsection{The local breaking of statistical isotropy}
In order to keep our model in the simplest, we consider the well-studied case of a massive vector field $A_\mu$ with a varying kinetic function \cite{Dimopoulos:2009am}
\begin{equation}\label{eq19}
{\cal L}_A=-\frac14\,f\,F_{\mu\nu}F^{\mu\nu}+\frac12\,m^2A_\mu A^\mu\,.
\end{equation}
Arguably, this is the simplest stable theory in which massive vector fields can be produced during inflation \cite{Himmetoglu:2008zp,Himmetoglu:2008hx,Dimopoulos:2009vu}. In order for the vector field to be substantially produced during inflation, the kinetic function of the vector field and its mass are allowed to have a time-dependence parametrized by
\begin{equation}\label{eq16}
f\propto a^\alpha\quad\textrm{and}\quad m\propto a^\beta\,.  
\end{equation}
Using this, a successful vector curvaton mechanism can be built, with a scale-invariant spectrum of vector perturbations for appropriate values\cite{Dimopoulos:2011ws} of $\alpha$ and $\beta$. 

In order to break statistical isotropy, we envisage the case when $f$ becomes modulated by the field $\sigma$ in Eq.~(\ref{eq10}), i.e. $f=f(\sigma)$. Of course, we also include the $\chi$ field sector and assume the necessary conditions so that $\sigma$ obtains a distribution of out-of-equilibrium patches at the end of inflation. In turn, the emergence of out-of-equilibrium patches, and hence the fact that in some regions $\sigma$ starts to oscillate before the end of inflation, entails that in such regions the scaling regime in Eq.~(\ref{eq16}) finishes before the end of inflation. As a result, the energy density of the vector field at the end of inflation depends on the spatial location, namely
\begin{equation}\label{eq17}
\rho_{A,{\rm end}}=\rho_{A,{\rm end}}(\mbox{\boldmath$x$})\,.
\end{equation}
In the next section, we demonstrate this spatial dependence by solving numerically the dynamics of the vector field. For now, we show how this spatial dependence is transfered to the curvature perturbation by means of the vector curvaton scenario \cite{Dimopoulos:2011ws}.

Following \cite{Dimopoulos:2009am}, the total curvature perturbation can be written as
\begin{equation}
\zeta=(1-\hat\Omega_A)\zeta_{\rm rad}+\hat\Omega_A\zeta_A\,,
\end{equation}
where $\zeta_{\rm rad}$ is the curvature perturbation present in the radiation dominated universe after inflation, $\zeta_A$ is the curvature perturbation in the vector field, $\hat\Omega_A\equiv3\Omega_A/(4-\Omega_A)$ sets the relative contribution of each component to the total curvature perturbation and $\Omega_A\equiv\rho_A/\rho$. Taking $\Omega_A\lesssim1$ when the vector curvaton decays, the anisotropic contribution to $\zeta$ can be approximated by 
\begin{equation}\label{eq14}
\zeta_{\rm ani}(\mbox{\boldmath$x$})=\hat\Omega_A\zeta_A\simeq\frac{\sqrt3}{4\pi}\left[\frac{(\Omega_A)_{\rm dec}}{(\Omega_A)_{\rm end}}\,\frac{H_*}{m_P}\right]\Omega_{A,{\rm end}}^{1/2}(\mbox{\boldmath$x$})\,.
\end{equation}
Arranging the end of the scaling regime for $f$ in out-of-equilibrium patches at the end of inflation, the factor in square brackets becomes independent of \mbox{\boldmath$x$}, but $\Omega_{A,{\rm end}}$ depends on \mbox{\boldmath$x$}, as suggested by Eq.~(\ref{eq17}). 

\subsection{Numerical computation}
For simplicity, we assume that the dynamics of the vector field does not backreact on $\sigma$ and follow a phenomenological approach to model the transition from the scaling regime $f\propto a^\alpha$ to the end of the scaling, where $f=1$. We model this transition by means of a time-dependent exponent $\alpha(t)$, varying from $\alpha=-4$ (corresponding to a scale-invariant vector perturbation spectrum) to $\alpha=0$, where $f=1$, in a lapse of a few $e$-foldings, as suggested by the numerical solution to the system in Eqs.~(\ref{eq11})-(\ref{eq13}).

However, the kinetic function is modulated by $\sigma$, which is distributed according to a probability density owing to its inflationary fluctuations. For concreteness, we take this probability to peak around a given field value when the largest observable scale crosses the horizon. Therefore, when the shortest CMB scale exits the horizon, the field's probability density, denoted by $P_\cmb(\sigma_0)$, is  a Gaussian with mean $\bar\sigma_0$ (which we treat as a free parameter) and variance $\Sigma^2_\cmb\simeq\frac{H^2}{4\pi^2}N_\cmb$, where $N_\cmb\simeq9$. Here, we stress that we are interested in the probability density that carries the information on $\sigma$-field correlations \emph{only} on CMB scales at the end of inflation. Consequently, we do not allow inflationary fluctuations to source $P_\cmb(\sigma_0)$ once CMB scales become superhorizon (see Refs.~\refcite{Sanchez:2014cya,Sanchez:2016lfw} for a detailed discussion), and hence $P_\cmb(\sigma_0)$ follows a deterministic evolution until it reaches the barrier. Once $\sigma=\sigma_c$ and a classical $\chi$ field starts becoming produced, fluctuations in the number density of produced $\chi$ particles will source fluctuations of the classical $\sigma$ (see for example Ref.~\refcite{Lee:2011fj}). Nevertheless, in the Hartree approximation the source term, proportional to $(\chi^2-\langle\chi^2\rangle)$, is neglected, and hence the production of the field $\chi$ does not induce further fluctuations of $\sigma$. Therefore, in the Hartree approximation the evolution of $P_\cmb(\sigma_0)$ remains fully deterministic until the end of inflation. Going beyond the Hartree approximation, one can expect that the additional fluctuations of $\sigma$ will increase the fluctuations in the energy density of the vector field at then end of inflation, thus contributing to $\zeta_{\rm ani}(\mbox{\boldmath$x$})$ in Eq.~(\ref{eq14}). Therefore, the Hartree approximation allows us to obtain the minimal effects regarding an statistically inhomogeneous anisotropic contribution to the total curvature perturbation, or the local breaking of statistical isotropy for short.  

\begin{figure}[htbp]
\centerline{\includegraphics[width=5in]{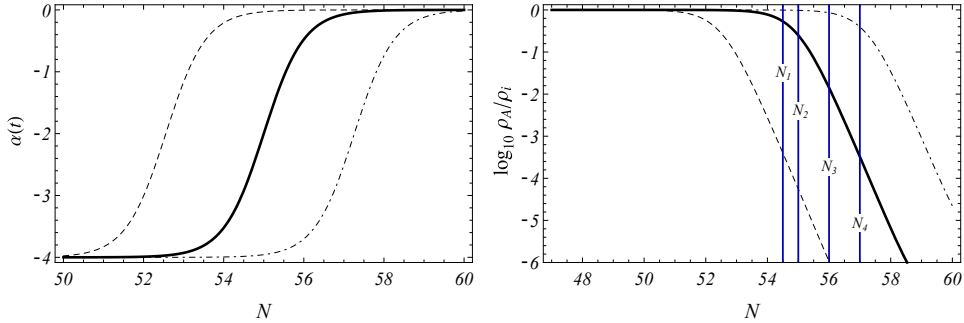}}
\caption{Evolution of $\alpha(t)$ (left panel) and energy density $\rho_A$ (right panel) during inflation.}\label{fig2}
\end{figure}

Taking all the above into account, different values of $\sigma$ (allowed by $P_\cmb(\sigma_0)$) imply that field interactions between $\sigma$ and $\chi$ become dynamically important at different times in different spatial locations. As previously discussed, in the Hartree approximation the exponent $\alpha(t)$ follows the same evolution in different regions but at different times, with the time delay determined by the amplitude of fluctuations of $\sigma$ present right after CMB scales become superhorizon. This is represented in Fig.~\ref{fig2}, where the thick solid line depicts the evolution of the exponent $\alpha(t)$ corresponding to the central value $\bar\sigma_0$, whereas dashed and dot-dashed lines depict the evolution corresponding to $\bar\sigma_0-3\Sigma_\cmb$ and $\bar\sigma_0-3\Sigma_\cmb$, respectively, thus implying an anticipated and delayed exit of the scaling regime for $f$.

Using the physical vector field $\mbox{\boldmath$W$}\equiv\sqrt f\mbox{\boldmath$A$}/a$ and its module $W=|\mbox{\boldmath$W$}|$, the evolution equation for the vector field reads
\begin{equation}
\ddot W+3H\dot W+\left[\frac14\frac{(\dot f)^2}{f^2}-\frac{\ddot f}{2f}-\frac{\dot f}{2f}\,H+2H^2+M^2\right]W=0\,,
\end{equation}
where $M^2=m^2/f$. Using the solution to the above we can compute the energy density of the vector field, determined by
\begin{equation}
\rho_A=\frac12H^2W^2\left(1-\frac12\frac{\dot f}{Hf}+\frac{\dot W}{HW}\right)^2+\frac12M^2W^2\,.
\end{equation}
In the righthand panel of Fig.~\ref{fig2} we plot the energy density, normalized with respect to its initial value $\rho_i$. We observe that as long as the scaling regime with $\alpha\simeq-4$ lasts, $\rho_A$ remains approximately constant, and starts decreasing when $\alpha\simeq0$. As a result of the different exit time of the scaling regime, the energy density of the vector field at the end of inflation becomes distributed according to a probability density. As we show in the next section, if inflation finishes at $N=N_3$ or $N=N_4$, a relatively small fraction of regions exist where $\rho_{A,{\rm end}}\simeq\rho_i$ whereas in other (more abundant) regions we find $\rho_{A,{\rm end}}\ll\rho_i$. This demonstrates both the dependence of $\Omega_A$ with position (and hence that of $\zeta_{\rm ani}(\mbox{\boldmath$x$})$) and the possibility of a local breaking of statistical isotropy.

\subsection{The curvature perturbation $\zeta_{\rm ani}(\mbox{\boldmath$x$})$ as a stochastic variable}
The purpose of this section is to transform the probability density $P_\cmb(\sigma_0)$, describing the distribution of $\sigma$ once all CMB scales cross outside the horizon, into a probability density where $\zeta_{\rm ani}(\mbox{\boldmath$x$})$ is the stochastic variable. This can be achieved by performing the necessary changes of variable. 

For a given spatial patch we denote by $t_c$ the associated critical time, defined by the condition $\sigma(t_c)=\sigma_c$. Since $\sigma$ is light enough to undergo particle production, the slow-roll approximation allows us to rewrite this condition as $\sigma_0=\sigma_c\,\exp\left(c_\sigma H t_c/3\right)$. Using the above and the conservation of probability, we may write express $P_\cmb(\sigma_0)$ in terms of the critical time as follows
\begin{equation}
P_\cmb(t_c)=\frac{d\sigma_0}{dt_c}\,P_\cmb(\sigma_0(t_c))\,.
\end{equation}

Given initial conditions $W_0,\dot W_0$ for $W$ and a critical time $t_c$, $\rho_A$ at the end of inflation is a function of the critical time $t_c$ only, i.e. $\rho_{A,{\rm end}}(t_c)\equiv\rho_A(N_{\rm end},t_c)$. Therefore, we can write $P_\cmb(t_c)$ as follows
\begin{equation}\label{eq15}
P_\cmb(\rho_{A,{\rm end}})=\frac{dt_c}{d\rho_{A,{\rm end}}}\,P_\cmb(t_c(\rho_{A,{\rm end}}))\,.
\end{equation}
To compute the above derivative we note that field histories with different $t_c$ are approximately the same after the proper synchronization. Using the history with $\bar\sigma_0$ as reference, we define the shift $\delta t\equiv t_c-\bar t_c$, where $\bar t_c$ is the critical time associated with $\bar\sigma_0$ (see Fig.~\ref{fig2}). Therefore, in the separate universe approach we have
\begin{equation}\label{eq4}
\rho_A(N_{\rm end},t_c)=\rho_A(N_{\rm end}-H\delta t,\bar t_c)\,,
\end{equation}
which affords us to compute $t_c=t_c(\rho_{A,{\rm end}})$, and hence the derivative in Eq.~(\ref{eq15}), from the numerical solution for $\rho_A$ depicted in the right-hand panel of Fig.~\ref{fig2}. Now, introducing the relative energy density $\rho_r\equiv\rho_{A,{\rm end}}/\rho_i$ we have
\begin{equation}
P_\cmb(\rho_r)=\frac{d\rho_{A,{\rm end}}}{d\rho_r}\,P_\cmb(\rho_{A,{\rm end}}(\rho_r))\,.
\end{equation}
In view of Eq.~(\ref{eq14}), we define the variable
\begin{equation}
\zeta_r(\mbox{\boldmath$x$})\equiv\zeta_{\rm ani}(\mbox{\boldmath$x$})/\zeta_i=\rho_r^{1/2}(\mbox{\boldmath$x$})\,,
\end{equation}
where $\zeta_i$ is the curvature perturbation contributed by the vector field when $\rho_{A,{\rm end}}\simeq\rho_i$. Or equivalently, when the scaling regime for $f$ lasts until the end of inflation. Using all the above we arrive at the desired result
\begin{equation}
P_\cmb(\zeta_r)=\frac{d\sigma_0}{dt_c}\,\frac{dt_c}{d\rho_{A,{\rm end}}}\,\frac{d\rho_{A,{\rm end}}}{d\rho_r}\,\frac{d\rho_r}{d\zeta_r}\,P_\cmb(\sigma_0(t_c(\rho_{A,{\rm end}}(\rho_r(\zeta_r)))))\,.
\end{equation}

\begin{figure}[htph]
\includegraphics[width=5in]{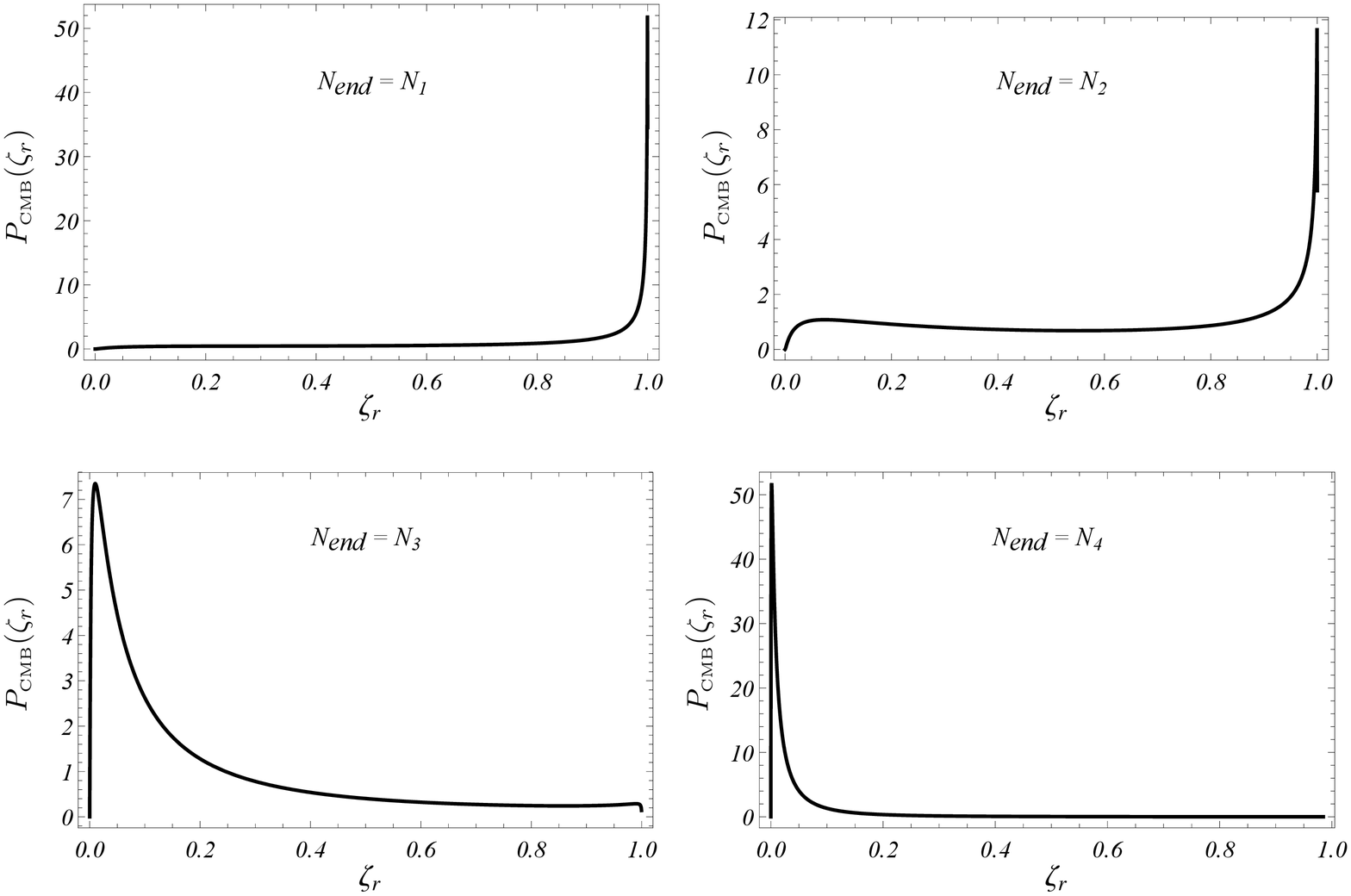}
\caption{Plot of the probability density $P_\cmb(\zeta_r)$ for different values of $N_{\rm end}$.}\label{fig3}
\end{figure}

In Fig.~\ref{fig3} we plot our results for the probability $P_\cmb(\zeta_r)$ when inflation finishes at different $N_{\rm end}$ (see also Fig.~\ref{fig2} for comparison). For $N_{\rm end}=N_1$, the probability $P_\cmb(\zeta_r)$ peaks around $\zeta_r=1$, indicating a statistically homogeneous contribution of the vector field to the curvature perturbation. In this case, a highly anisotropic vector perturbation becomes incompatible with observations unless it gives a subdominant contribution to the total curvature perturbation \cite{Dimopoulos:2009am,Dimopoulos:2009vu,Dimopoulos:2011ws}. For $N=N_2$, $P_\cmb(\zeta_r)$ still accumulates predominantly around $\zeta_r=1$, but in some regions, making up a non-negligible fraction of the observable Universe, the scaling regime for $f$ is already over. Therefore, in those regions we find $\zeta_{\rm ani}<\zeta_i$. Consequently, for $N\geq N_2$ the vector field perturbation starts becoming statistically inhomogeneous. For $N=N_3$, we find that only in a limited fraction of the observable Universe it becomes possible to find $\zeta_{\rm ani}\simeq\zeta_i$, whereas in the remaining part $\zeta_{\rm ani}<\zeta_i$. Tuning the parameters of the vector field model, one can arrange that $\zeta_i\sim10^{-5}$, and hence the contribution of the vector field becomes observable in out-of-equilibrium patches only. Since the vector perturbation is statistically very inhomogeneous, the CMB anisotropy constraint on cosmological vector fields \cite{Hanson:2010gu,Ramazanov:2013wea,Kim:2013gka,Ade:2015hxq,Ade:2015lrj}, obtained under the assumption of statistical homogeneity of the perturbation, can be consequently evaded. As a result, the vector field may be allowed to imprint a highly anisotropic contribution to the curvature perturbation. This argument has been recently put forward in Ref.~\refcite{Sanchez:2016lfw} to hypothesize the existence of a vector spot in the CMB. Finally, for $N=N_4$ the curvature perturbation $\zeta_{\rm ani}$ becomes too small to be observable.

\section{Conclusions}
In this talk we follow an approach recently presented in the literature, aiming to provide a common framework for the origin of the large-angle CMB anomalies, to elaborate in detail on the breaking of statistical isotropy employing a vector field model in which the kinetic function $f$ is modulated by an isocurvature field $\sigma$ that can fluctuate during inflation. Thanks to a preexisting stage of fast-roll inflation, $\sigma$ features an out-of-equilibrium configuration at the beginning of the slow-roll, but due to its coupling to the scalar sector $\chi$, at the end of inflation such an out-of-equilibrium configuration remains only in sparse regions of the observable Universe. 

To investigate the implications of the patchy structure of $\sigma$ for the vector field, we study a phenomenological model to account for the variation of $f$ induced by the dynamics of the isocurvature field $\sigma$, which we solve numerically assuming no backreaction from the vector field. Motivated by the dynamics of $\sigma$, in the proposed model we impose that the transition from the scaling regime with $\alpha=-4$ (allowing the inflationary production of the vector field) to $f=1$ (causing the rapid depletion of the vector field's energy density) takes place in a few $e$-foldings (see Fig.~\ref{fig2}). This transition is triggered at the critical time $t_c$, that depends on the field value $\sigma_0$ (taken after CMB scales become superhorizon). Therefore, using this model and a particular $t_c$ (corresponding to a given spatial patch), we solve numerically the evolution of the vector field $W$ and its energy density $\rho_A$. However, to compute $\rho_{A,{\rm end}}$ in different regions of the Universe we must take into account the fluctuations obtained by $\sigma$ while CMB scales exit the horizon, encoded in the probability density $P_\cmb(\sigma_0)$. Such fluctuations are transferred to $f$, thus giving rise to fluctuations in $\rho_{A,{\rm end}}$. Our results show that if the scaling regime finishes during inflation, the fluctuations in $\rho_A$ can span several orders of magnitude (see Fig.~\ref{fig2}).

Finally, we examine the implications of the fluctuations in $\rho_{A,{\rm end}}$ for the curvature perturbation contributed by the vector field $\zeta_{\rm ani}$ (see Eq.~(\ref{eq14})). To do so, we transform the probability density $P_\cmb(\sigma_0)$ into a probability density where the stochastic variable $\zeta_{\rm ani}$, which takes different values in different regions. In Fig.~\ref{fig3} we plot the resulting probability $P_\cmb(\zeta_r)$, where $\zeta_r\propto \zeta_{\rm ani}$, for different situations at the end of inflation (compare with Fig.~\ref{fig2}). Our results, although derived from a phenomenological model, demonstrate that the large fluctuations in $\rho_{A,{\rm end}}$ give rise to a statistical inhomogeneous contribution $\zeta_{\rm ani}(\mbox{\boldmath$x$})$ which, after tuning the parameters of the vector field model, can be arranged to be observable only in sparse regions of the Universe, thus leading to a local breaking of statistical isotropy. 

\section*{Acknowledgments}
The author thanks Juan P. Beltr\'an Almeida for reviewing the manuscript. The author is supported by COLCIENCIAS grant No. 110656399958.

\bibliographystyle{h-physrev}
\bibliography{references}

\end{document}